\documentstyle[12pt]{article}
\title{Some remarks on a nongeometrical interpretation of gravity and the 
  flatness problem}
\author{Hrvoje Nikoli\'c  \\
Theoretical Physics Division, Rudjer Bo\v{s}kovi\'{c} Institute, \\
P.O.B. 1016, HR-10001 Zagreb, Croatia \\
{\normalsize hrvoje@faust.irb.hr} \\
\makebox[1in]{} \\
}
\date{\today}
\begin{document}
\maketitle
\begin{abstract}
In a nongeometrical interpretation of gravity, the metric 
$g_{\mu\nu}(x)=\eta_{\mu\nu}+\Phi_{\mu\nu}(x)$ is interpreted as an 
{\em effective} metric, whereas $\Phi_{\mu\nu}(x)$ is interpreted as 
a fundamental gravitational field, propagated in spacetime which is 
actually flat. 
Some advantages and disadvantages of such an interpretation 
are discussed. The main advantage is a natural resolution 
of the flatness problem.   
\end{abstract}
\vspace*{0.5cm}

\section{Introduction}

It seems that a gravitational theory based on a scalar or a vector field 
in a flat Minkowski space cannot describe known experimental data
\cite{feyn}, \cite{mtw}. On the other hand, the fenomenological success of
Einstein's theory of gravity suggests that gravity should be described 
completely, or at least partially, by a symmetric second-rank tensor field. In 
general, a symmetric second-rank tensor field contains components of 
spin-0, spin-1 and spin-2 \cite{barn}. There are many theories of gravity 
based on a symmetric second-rank tensor field \cite{fuchs}, \cite{zeld}. 
However, if we require that 
a symmetric second-rank tensor $\Phi_{\mu\nu}$ describes a massless spin-2 
field in a flat Minkowski space with metric $\eta_{\mu\nu}$ and satisfies 
a second-order differential 
equation in which $\Phi_{\mu\nu}$ is consistently coupled to itself and to 
other fields, then the most general such equation can be written in the form 
of the Einstein equation (with a cosmological term)  
\cite{feyn2}-\cite{gris}, where the ``effective metric" is
given by  
\begin{equation}\label{efect}
g_{\mu\nu}(x)=\eta_{\mu\nu}+\Phi_{\mu\nu}(x) \; .
\end{equation} 
The Einstein equation, when written in terms of
$\Phi_{\mu\nu}$ and $\eta_{\mu\nu}$, possesses an infinite number of terms. 
On the other hand, this equation looks much simpler when it is written in 
terms of $g_{\mu\nu}$. This suggests, but in no way proves, that
$g_{\mu\nu}$, and not $\Phi_{\mu\nu}$, is a fundamental field. Such an 
interpretation leads 
to the standard geometrical interpretation of gravity. However, such an 
interpretation makes gravity very different from other fields, because 
other fields describe some dynamics for which spacetime serves as a
background, while gravity describes the dynamics of spacetime 
itself. This may be one of the obstacles to formulate a consistent 
theory of quantum gravity. 

The aim of this paper is to investigate a 
nongeometrical interpretation (NGI) of gravity, in which $\Phi_{\mu\nu}(x)$ is a 
fundamental gravitational field propagated in a flat Minkowski spacetime  
with the metric $\eta_{\mu\nu}$, while $g_{\mu\nu}(x)$ has the 
role of the effective 
metric only. Some aspects of such an interpretation have already been
discussed  
\cite{thir}. In this paper we reconsider some conclusions drawn in \cite{thir} 
and stress some novel conclusions.  
We find that such an interpretation is not only
consistent, but also leads to several advantages with respect to the 
standard interpretation. In particular, it leads to 
a natural resolution of the flatness problem. We also comment on 
some disadvantages of such an interpretation. 

\section{Global topology and cosmology in the NGI} 

It has recently been suggested \cite{nikolic} that gravity, as a dynamical 
theory of the metric tensor $g_{\mu\nu}(x)$, should not be interpreted as a 
dynamical theory of the space-time topology. The topology should be 
rather fixed by an independent axiom, while the Einstein (or some other) 
equation determines only the metric tensor on a fixed manifold. For the  
Cauchy problem to be well posed, it is neccessary that the topology is of the 
form $\Sigma\times{\bf R}$. The most natural choice is ${\bf R}^D$ as a 
global topology, which admits a flat metric $\eta_{\mu\nu}$. Thus the 
NGI of gravity, which we consider in this paper, 
supports this nontopological interpretation, because in the NGI 
it is manifest that the topology is fixed by the background 
spacetime with a flat metric $\eta_{\mu\nu}$.  

The nongeometrical (or nontopological) interpretation may seem to 
be inconsistent on global level, because it
starts with a global ${\bf R}^D$ topology of spacetime, while the Einstein 
equation, which determines $\Phi_{\mu\nu}$ and $g_{\mu\nu}$, possesses  
solutions for the metric $g_{\mu\nu}$ which correspond to a different
topology. 

However, this problem is resolved in the Cauchy-problem approach. 
For example, if the space has ${\bf R}^3$ topology on the ``initial" Cauchy
surface, then it has the same topology at all other instants. Quite
generally, if the Cauchy problem is well posed, then the space topology
cannot change during the time evolution \cite{konst}. The fact that the topology 
of time in the Friedman universes is not ${\bf R}$, but a connected 
submanifold of ${\bf R}$ which is singular on its end(s), 
can be interpreted merely as a sign of nonaplicability of 
the Einstein equation for high-energy densities. 

However, the interesting question is whether the 
NGI is consistent if the  
Einstein equation is not treated as a Cauchy problem and 
singularities are not treated as pathologies of the model. In \cite{thir} 
it was concluded that the NGI of gravity was not 
appropriate for cosmological problems. Contrary to this conclusion, we argue 
that the application of the NGI of gravity to 
cosmological problems is actually the main advantage of this interpretation 
with respect to the conventional interpretation, because the NGI predicts that 
the effective metric $g_{\mu\nu}$ of a homogeneous and isotropic universe is 
flat, in agreement with observation. In the conventional approach, the 
assumption that the Universe is homogeneous and isotropic leads to the 
Robertson-Walker metric
\begin{equation}\label{robwalk}
 ds^2 =dt^2 -R^2(t)\frac{dx^2 +dy^2 +dz^2}{\left[1+(k/4) 
 (x^2 +y^2 +z^2)\right] ^2 } \; .
\end{equation}
If $k=0$, this corresponds to a flat universe. The observed flatness cannot 
be explained in the conventional approach. However, in the NGI,  
(\ref{robwalk}) is interpreted as an effective metric, whereas the  
fundamental quantity is the gravitational field $\Phi_{\mu\nu}$. The 
nonvanishing components of $\Phi_{\mu\nu}$ in (\ref{robwalk}) are 
\begin{equation}\label{comp}
 \Phi_{ij}(x)=\left\{ 1- \frac{R^2(t)}{\left[1+(k/4)
 (x^2 +y^2 +z^2)\right] ^2 } \right\} \delta_{ij} \; , \;\;\; 
  i,j=1,2,3 \; .
\end{equation}
Now the assumption that the Universe is homogeneous and isotropic means 
that $\Phi_{\mu\nu}$ does not depend on $x,y,z$, which leads to the
conclusion that the relation $k=0$ {\em must} be satisfied.   

\section{The question of local consistency of the NGI}

The fact that the NGI leads to a natural 
resolution of the flatness problem suggests that the NGI could be the right 
interpretation. Thus, it is worthwhile to further explore the consistency of 
such an interpretation. 

Let us start with the motion of a particle in a gravitational field. 
If we neglect the contribution of the particle to the gravitational field 
$\Phi_{\mu\nu}(x)$, then the action of the particle with a mass $m$ can be
chosen to be \cite{feyn3}, \cite{mtw}
\begin{equation}\label{act}
 S=m\left[ -\frac{1}{2}\int d\tau \: \dot{x}^{\mu}\dot{x}^{\nu}\eta_{\mu\nu} 
  -\kappa\int d\tau \: h_{\mu\nu}(x) \dot{x}^{\mu}\dot{x}^{\nu} \right] \; ,
\end{equation}
where $\tau$ is the proper time of the particle, $\dot{x}^{\mu}=dx^{\mu}/d\tau$, 
$h_{\mu\nu}(x)$ is a redefined gravitational field $2\kappa h_{\mu\nu}(x) 
\equiv \Phi_{\mu\nu}(x)$, and $\kappa$ is a coupling constant. The value 
of $\kappa$ is determined by the definition of $h_{\mu\nu}(x)$. For example,
$h_{\mu\nu}(x)$ can be defined such that, in the weak-field limit,  
$h_{00}(x)$ is equal to Newton's gravitational potential. The action (\ref{act}) 
also can be written as 
\begin{equation}\label{act1}
 S=m\left[ -\frac{1}{2}\int d\tau \: g_{\mu\nu}(x)\dot{x}^{\mu}\dot{x}^{\nu}
  \right] \; ,
\end{equation}
which is the convential form of the action of the particle in
the gravitational field. Both forms of the action lead to the same equations  
of motion which determine the trajectory $x^{\mu}(\tau )$. In the 
conventional geometrical interpretation, this trajectory is 
interpreted as a motion along a geodesic, which is not the case for the 
NGI. 

In (\ref{act}) and (\ref{act1}) it was stated that $\tau$ is the proper time,
but the proper time was not defined. For (\ref{act}) one could naively 
take the definition $d\tau^2 = \eta_{\mu\nu}dx^{\mu}dx^{\nu}$. On the 
other hand, in (\ref{act1}) the proper time is 
defined as $d\tau^2 = g_{\mu\nu}dx^{\mu}dx^{\nu}$, which leads to results  
which are in agreement with observations. 
We require that  
(\ref{act}) is equivalent to (\ref{act1}), so in (\ref{act}) we must take 
\begin{equation}\label{tau}
 d\tau^2 =[\eta_{\mu\nu} + 2\kappa h_{\mu\nu}(x)]dx^{\mu}dx^{\nu} \; .  
\end{equation} 

It is interesting to note that the existence of the geometrical 
interpretation is in no way the property of the symmetric second-rank 
tensor field only. For example, as noted in \cite{thir}, the interaction 
of a particle with a scalar field $\phi (x)$ can be described 
by the interaction part of the action $S_I =-m\kappa \int d\tau \: 
\phi (x)\dot{x}^{\mu}\dot{x}^{\nu}\eta_{\mu\nu}$, which leads to the 
action of the form (\ref{act1}), where the effective metric is 
$g_{\mu\nu}(x)=\eta_{\mu\nu}(1+2\kappa\phi (x))$.     

Now a few comments on the interpretation of various components of
$g_{\mu\nu}$. For example, if $g_{00}$ depends on $x$, in the conventional 
interpretation this is interpreted as a phenomenon that the lapse  
of time depends on $x$. In the NGI,  
it is interpreted that the effect of gravity is such that all kinds of
matter (massive and massless) move slower or faster, depending on $x$. 
Because of the equivalence principle (the coupling constant $\kappa$ in 
(\ref{act}) is the same for all kinds of particles), the motion of all 
kinds of matter is changed in the same way, 
namely, in such a way as if  
the metric of the time itself depended on $x$. Similarly, if $g_{ij}$ 
depends on $x$, in the NGI it is interpreted that 
the effect of gravity is such that all kinds of matter are contracted or 
elongated in the same way, depending on $x$. More details 
on this aspect of the NGI can be found in \cite{thir}.                   

In the NGI, the actual distances are given by $\eta_{\mu\nu}$ instead of 
by $g_{\mu\nu}$. For example, 
the actual time distance is given by $dt$ instead of by $\sqrt{g_{00}}dt$. 
Similarly, the actual space distance in the $x^1$-direction is given by $dx^1$  
instead of by $\sqrt{g_{11}}dx^1$. Consequently, the actual velocity 
of light $d{\bf x}/dt$ (with $ds^2=0$) is no longer a constant. However,    
as stressed in \cite{thir}, these actual distances 
are unobservable. Only the effective metric $g_{\mu\nu}$ can 
be measured. This is one of the unpleasent features of the NGI, but this  
does not make it inconsistent. 

However, there is even a more serious problem of 
the NGI. These actual distances are not only unobservable, but 
they are not uniquely defined, because of the 
invariance with respect to general coordinate transformations of 
the Einstein equation.  
The NGI makes sense only if some coordinate condition is fixed. 
If we can somehow find the right coordinate condition, then we can also 
define the actual distances. However, it is difficult to find this, because   
all coordinate conditions lead to the same observable effects, at least 
in classical physics. 

However, it is possible that, in quantum gravity, different coordinate
conditions are not equivalent. Moreover, some alternative classical theories 
of gravity do not possess the invariance with respect to general coordinate
transformations (see, for example, \cite{log}). All this suggests that,  
perhaps, there is a possibility, at least in principle, of identifying   
the right coordinate condition experimentally. At present, we can only guess 
what that might be, using some simplicity and symmetry arguments. 
If we require that this condition should be expressed in terms of $\eta_{\mu\nu}$ 
and $\Phi_{\mu\nu}$, and that this should not violate Lorentz covariance, 
then the simplest choice is the harmonic condition 
\begin{equation}\label{harm}
 D^{\mu}\Phi_{\mu\nu}=0 \; ,
\end{equation}
where $D^{\mu}$ is the covariant derivative with respect to a {\em flat} 
metric (i.e., a metric which can be transformed to $\eta_{\mu\nu}$ by a 
coordinate transformation). This condition is preferred by many 
authors \cite{krai}, \cite{thir}. The metric (\ref{robwalk}) does not satisfy 
this condition, but one can easily transform (\ref{robwalk}) 
into coordinates for which this condition is satisfied, and conclude 
in the same way that $k=0$. One can also see that (\ref{comp}) for $k=0$ 
already satisfies (\ref{harm}). 

\section{Conclusion}   

The NGI of gravity is consistent and leads to a  
natural resolution of the flatness problem. The flatness problem 
can also be resolved by the inflationary model, which predicts that 
today the Universe should be very close to be flat, 
even if it was not so flat in early stages of its evolution. 
On the other hand, the NGI 
predicts that in a homogeneous and isotropic universe, the {\em exact} 
flatness must be observed in {\em all} stages of its evolution. 
Both predictions are in agreement with present
observational data.   
 
The gravitational 
field $\Phi_{\mu\nu}(x)$ does not differ much from other fields, 
because it is a field propagated in a nondynamical flat spacetime. 
The consistency of the NGI requires that some 
coordinate condition should be fixed, so the resulting theory is no longer 
covariant with respect to general coordinate transformations. 
However, the Einstein equation written in terms of $\eta_{\mu\nu}$ and 
$\Phi_{\mu\nu}$, and supplemented by (\ref{harm}), is Lorentz covariant. 

The disadvantages of the NGI are the following: The actual metric
$\eta_{\mu\nu}$ is 
unobservable, only the effective metric $g_{\mu\nu}$ can be measured, 
at least if the equivalence principle is exact.  
The Einstein equation seems very complicated when written in terms 
of $\Phi_{\mu\nu}(x)$ and $\eta_{\mu\nu}$. The action for a particle 
in a gravitational field, given by (\ref{act}) and (\ref{tau}), 
in the NGI also seems more complicated than in the conventional,  
geometrical interpretation. 

However, if some of the 
alternative theories of gravity is more appropriate than the theory 
based on the Einstein equation, it is possible that the equivalence
principle is not exact and that the correct equation of motion is 
not so complicated when written in terms of $\eta_{\mu\nu}$,
$\Phi_{\mu\nu}(x)$, and possibly some additional dynamical fields. 
 
\section*{Acknowledgment}

The author is grateful to N. Bili\'{c} and H. \v{S}tefan\v{c}i\'{c} 
for some useful suggestions. 
This work was supported by the Ministry of Science and Technology of the
Republic of Croatia under Contract No. 00980102.

\end{document}